\documentclass[11pt,a4paper,twoside,fleqn]{article}
\usepackage{color,graphicx,natbib,lscape,here,geometry,setspace,multirow,enumitem}
\usepackage[dvipsnames]{xcolor}
\usepackage{graphicx}
\usepackage{amsmath,amssymb,bm}

\usepackage{authblk}
\usepackage{silence}
\usepackage{rotating}
\usepackage{multirow}

\usepackage{booktabs}
\usepackage{diagbox}
\usepackage{color}
\usepackage{setspace}
\usepackage{algorithm2e}
\usepackage{enumitem}
\usepackage{tikz}

\definecolor{nblue}{HTML}{000660}
\usepackage[colorlinks=true,urlcolor=nblue,linkcolor=nblue,citecolor=nblue]{hyperref}
\usepackage{bigstrut}

\usepackage{tabularx}
\usepackage{threeparttable}
\usepackage{dcolumn}
\newcolumntype{d}[1]{D{.}{.}{#1}}

\usepackage{etoolbox} 
\makeatletter
\patchcmd{\BR@backref}{\newblock}{\newblock[}{}{}
\patchcmd{\BR@backref}{\par}{]\par}{}{}
\makeatother

\usepackage{pdflscape}
\usepackage{afterpage}

\usepackage{longtable}
\usepackage{array}
\newcolumntype{C}[1]{>{\centering\arraybackslash}p{#1}}

\usepackage{comment}
\usepackage{mathtools}

\usepackage[hang,flushmargin]{footmisc} 

\usepackage[hang]{footmisc}
\usepackage[british]{babel}

\pdfoutput=1
 \usepackage{float}
 \restylefloat{table}

\usepackage[title,titletoc]{appendix}
\makeatletter

\renewenvironment{appendices}{%
    \begin{oldappendices}%
    \renewcommand{\thefigure}{\ifnum \c@section>\z@ \thesection.\fi\@arabic\c@figure}%
    \@addtoreset{figure}{section}%
    \renewcommand{\thetable}{\ifnum \c@section>\z@ \thesection.\fi\@arabic\c@table}%
    \@addtoreset{table}{section}}{%
    \end{oldappendices}%
}\makeatother

\usepackage{titlesec} 
\titleformat{\section}[block]{\large}{\thesection. }{0em}{\MakeUppercase} 
\titleformat{\subsection}[block]{\large}{\thesubsection. }{0em}{\itshape} 
\titleformat{\subsubsection}[block]{\large}{}{0em}{\itshape} 

\let\natbibcitet\citet
\renewcommand\citet{\bibpunct{(}{)}{,}{a}{,}{,}\natbibcitet}
\let\natbibcitep\citep
\renewcommand\citep{\bibpunct{(}{)}{;}{a}{,}{;}\natbibcitep}
\newcommand{\bi}{\begin{itemize}}
\newcommand{\ei}{\end{itemize}}
\newcommand{\be}{\begin{equation}}
\newcommand{\ee}{\end{equation}}
\defcitealias{ieo14}{IEO, 2014}

\long\def\symbolfootnote[#1]#2{\begingroup%
\def\thefootnote{\fnsymbol{footnote}}\footnote[#1]{#2}\endgroup}

\makeatletter
\def\ubar#1{\underline{\sbox\tw@{$#1$}\dp\tw@\z@\box\tw@}}
\def\obar#1{\overline{\sbox\tw@{$#1$}\dp\tw@\z@\box\tw@}}
\makeatother

\usepackage{bm}
\usepackage{caption}
\usepackage{subcaption}

\makeatletter\let\p@subfigure\thefigure\makeatother

\floatstyle{plaintop}
\restylefloat{table}

\usepackage{fancyhdr} 
\usepackage{cleveref}
\crefname{chapter}{Chapter}{Chapters}
\crefname{section}{Section}{Sections}
\crefname{subsection}{Section}{Sections}
\crefname{subsubsection}{Section}{Sections}
\crefname{figure}{Figure}{Figures}
\crefname{table}{Table}{Tables}
\crefname{equation}{Equation}{Equations}
\crefname{appendix}{Appendix}{Appendices}
\crefname{appendices}{Appendix}{Appendices}

\crefname{appsec}{Appendix}{Appendices}

\usepackage{catoptions}
\makeatletter

\def\Autoref#1{%
  \begingroup
  \edef\reserved@a{\cpttrimspaces{#1}}%
  \ifcsndefTF{r@#1}{%
    \xaftercsname{\expandafter\testreftype\@fourthoffive}
      {r@\reserved@a}.\\{#1}%
  }{%
    \ref{#1}%
  }%
  \endgroup
}
\def\testreftype#1.#2\\#3{%
  \ifcsndefTF{#1autorefname}{%
    \def\reserved@a##1##2\@nil{%
      \uppercase{\def\ref@name{##1}}%
      \csn@edef{#1autorefname}{\ref@name##2}%
      \autoref{#3}%
    }%
    \reserved@a#1\@nil
  }{%
    \autoref{#3}%
  }%
}
\makeatother

\title{\LARGE{Measuring international uncertainty using global vector autoregressions with drifting parameters}}
\author{\large{\uppercase{Michael Pfarrhofer}}\thanks{
\noindent Salzburg Centre of European Union Studies, University of Salzburg. \textit{Address}: M\"{o}nchsberg 2a, 5020 Salzburg, Austria. \textit{Email}: \href{mailto:michael.pfarrhofer@sbg.ac.at}{michael.pfarrhofer@sbg.ac.at}. \textit{Phone}: +43 662 8044 3772. The author thanks Martin Feldkircher, Sylvia Fr\"uhwirth-Schnatter, Niko Hauzenberger, Florian Huber, Gregor Kastner and Anna Stelzer for valuable comments and suggestions. Funding from the Austrian Science Fund (FWF) for the project ``High-dimensional statistical learning: New methods to advance economic and sustainability policies'' (ZK 35) is gratefully acknowledged.}
\\\vspace*{-0.5em}
\textit{University of Salzburg}}
\date{}


\def\equationautorefname~#1\null{%
  Eq.~(#1)\null
}
\def\equationautorefname~#1\null{
Eq.~(#1)\null
}

\usepackage[utf8, latin1]{inputenc}                 

\begin{document}
\maketitle\thispagestyle{empty}\normalsize\vspace*{-2em}\small

\begin{center}
\begin{minipage}{0.8\textwidth}
\noindent\small This paper investigates the time-varying impacts of international macroeconomic uncertainty shocks. We use a global vector autoregressive specification with drifting coefficients and factor stochastic volatility in the errors to model six economies jointly. The measure of uncertainty is constructed endogenously by estimating a scalar driving the innovation variances of the latent factors, which is also included in the mean of the process. To achieve regularization, we use Bayesian techniques for estimation, and introduce a set of hierarchical global-local priors. The adopted priors center the model on a constant parameter specification with homoscedastic errors, but allow for time-variation if suggested by likelihood information. Moreover, we assume coefficients across economies to be similar, but provide sufficient flexibility via the hierarchical prior for country-specific idiosyncrasies. The results point towards pronounced real and financial effects of uncertainty shocks in all countries, with differences across economies and over time.
\end{minipage}
\end{center}

\onehalfspacing\normalsize\renewcommand{\thepage}{\arabic{page}}
\newpage\setcounter{page}{1}

\section{Introduction}\label{sec:introduction}
Uncertainty has received a substantial amount of attention as a driving force of business cycle fluctuations, following the experiences of economists and policy makers in the aftermath of the Great Recession. Measuring uncertainty and its impact on the economy produced a voluminous literature, with prominent contributions including \citet{doi:10.3982/ECTA6248,jurado2015measuring,doi:10.1093/qje/qjw024,doi:10.3982/ECTA13960,carriero2016measuring}, among many others. 

These studies provide compelling theoretical and empirical evidence suggesting negative economic consequences of uncertainty shocks. Elevated levels of uncertainty can produce large drops in economic activity, and moreover render counteracting monetary and fiscal policies less effective \citep[see, e.g.][]{aastveit2013economic,doi:10.1002/jae.2723}. Transmission channels of uncertainty shocks to the macroeconomy relate mainly to real phenomena such as distorted corporate decision making, while recent papers highlight the importance of disturbances on credit and financial markets \citep{doi:10.3982/ECTA6248,ALESSANDRI2018}. 

The econometric literature increasingly relies on a unified framework for estimating uncertainty and its effects jointly \citep[see][]{mumtaz2013impact,carriero2015impact,carriero2016measuring,doi:10.1002/jae.2613}. Besides many contributions relying on linear specifications assuming model parameters other than the residual variances to be constant, there has been a recent focus on accounting for nonlinear relationships between uncertainty and the real and financial economy, with potential consequences for the measurement of uncertainty \citep{mumtaz2017changing,ALESSANDRI2018,mumtaz2019evolving}. In addition, given the importance of international linkages in the propagation of macroeconomic shocks \citep[see, e.g.][]{CANOVA2004327,Pesaran2004,mumtaz2009transmission,FELDKIRCHER2016167}, multi-economy modeling frameworks have been proposed in \citet{mumtaz2015international,berger2016global,crespo2017macroeconomic,carriero2018assessing} and \citet{mumtaz2019evolving}. A theoretical justification for empirically assessing spillover dynamics is provided in \citet{mumtaz2017common}, who identify increasing globalization and trade openness as main determinant of international volatility comovements. Note, however, that most of these contributions rely on factor models or focus on specific variables, rather than providing a full systematic procedure addressing both the real and financial economic sectors across countries.

While some papers either consider time-variation in the relationship between uncertainty and the macroeconomy, or country-specific idiosyncrasies stemming from different domestic dynamics or spillover effects, papers addressing both features jointly are limited \citep[with the exception of][]{mumtaz2019evolving}. The focus in \citet{mumtaz2019evolving}, however, is on the measurement of international, regional and country-specific uncertainty factors and their contribution to macroeconomic fluctuations over the cross-section, and not directly on structural inference based on uncertainty shocks for a high-dimensional system of multiple economies. Motivated by these notions, we propose a multi-economy model with drifting coefficients and factor stochastic volatility in mean to estimate uncertainty and its effects jointly. Allowing the volatilites to affect both the first and second moments of the multivariate dynamic system relates to stochastic volatility in mean models \citep[see][]{koopman2002stochastic,chan2017stochastic}.

We provide two empirical and econometric contributions: First, we extend the global vector autoregressive model of \citet{Pesaran2004} to feature time-varying parameters and residual variances that also enter the mean of the process \citep[for a related approach, see][]{doi:10.1111/rssa.12439}. Second, though this modeling framework decreases the number of parameters compared to unrestricted estimation, the parameter space is still high-dimensional. As a remedy, we employ Bayesian methods and adapt global-local priors for state-space models \citep[see][]{FRUHWIRTHSCHNATTER201085,doi:10.1002/for.2276,bittosfs,huber2019inducing} in a multi-country context. We center the model on a constant parameter specification with homoscedastic errors and cross-country homogeneity, but preserve the possibility of time-variation and heterogeneous dynamics across economies via flexible hierarchical priors. The hierarchical prior setup is designed to efficiently exploit cross-sectional information for precise inference, and relates to the Bayesian treatment of panel data \citep{verbeke1996linear,fruhwirth2004bayesian}.

Our model is applied to monthly data for six economies (France, Germany, Italy, Japan, the United Kingdom, and the United States) for the period ranging from 1991:04 to 2018:07, providing a link to the contributions of \citet{crespo2017macroeconomic} and \citet{carriero2018assessing} while generalizing the setup to allow for nonlinear relationships between uncertainty and the economy. The information set includes several recessionary episodes, and thus periods of economic distress when uncertainty is typically perceived to play a major role. The endogenous measure of uncertainty is comparable to established proxies, and links well to events associated with high uncertainty. Impulse responses shed light on the effects of uncertainty shocks on a set of macroeconomic and financial quantities. The responses are in line with established findings, but differ across the six countries in terms of magnitude and timing. Some variables show systematic declines in their responsiveness to uncertainty shocks, while others remain comparatively stable, corroborating findings in \citet{mumtaz2017changing}. For selected quantities in a subset of economies, the time-varying effects of uncertainty shocks do not evolve gradually, but exhibit distinct features for specific periods \citep[similar to][]{ALESSANDRI2018}.

The article is organized as follows. Section \ref{sec:econometrics} proposes the global vector autoregressive model with drifting coefficients and factor stochastic volatility in mean. Section \ref{sec:data} presents the data and discusses model specification. Section \ref{sec:results} investigates the uncertainty measure and provides a discussion of the empirical results. Section \ref{sec:conclusions} concludes.

\section{Econometric framework}\label{sec:econometrics}
\subsection{Model specification}
Let $\bm{y}_{it}$ denote a $k\times1$ vector of endogenous variables for $t=1,\hdots,T$ specific to country $i=1,\hdots,N$. Collecting country-specific endogenous variables yields the $K\times1$ vector $\bm{y}_{t}=(\bm{y}_{1t}',\hdots,\bm{y}_{Nt}')$ with $K=kN$, while we stack the reduced form shocks to $\bm{y}_{it}$ in a $K\times1$ vector $\bm{\epsilon}_{t} = (\bm{\epsilon}_{1t}',\hdots,\bm{\epsilon}_{Nt}')'$. 

Following \citet{Aguilar00bayesiandynamic}, we consider a factor stochastic volatility structure on the error term,
\begin{equation}
\bm{\epsilon}_{t} = \bm{L} \bm{f}_t + \bm{\eta}_{t}, \quad \bm{f}_t \sim \mathcal{N}(\bm{0},\exp(h_t)\times\bm{\Sigma}), \quad \bm{\eta}_{t}\sim\mathcal{N}(\bm{0},\bm{\Omega}_t).\label{eq:error}
\end{equation}
Here, $\bm{f}_t$ is a vector of $d\times1$ common static factors (with $d \ll K$), and $\bm{\eta}_{t}$ an idiosyncratic white noise shock vector of dimension $K\times1$. Latent factors are linked to the errors by the $K\times d$ factor loadings matrix $\bm{L}$. The factors $\bm{f}_t$ are Gaussian with zero mean and common time-varying volatility $\exp(h_t)$, scaling a diagonal matrix $\bm{\Sigma} = \bm{I}_d$, with $\bm{I}_{d}$ referring to a $d$-dimensional identity matrix.

The idiosyncratic error components $\bm{\eta}_{t}$ are assumed to follow a Gaussian distribution centered on zero with $K\times K$ time-varying covariance matrix
\begin{equation*}
\bm{\Omega}_t = \text{diag}(\exp(\omega_{1t}),\hdots,\exp(\omega_{Kt}))
\end{equation*} 
For both the volatility of the factors and the variances of the idiosyncratic components, we rely on a stochastic volatility model \citep{10.2307/1392151}. Here, $h_t$ and $\omega_{ij,t}$ for $i=1,\hdots,N$ and $j=1,\hdots,k$ follow independent random walk processes
\begin{align*}
&h_{t} = h_{t-1} + \xi_{t}, \quad \xi_{t}\sim\mathcal{N}(0,\sigma_{h})\\
&\omega_{ij,t} = \omega_{ij,t-1} + \zeta_{t}, \quad \zeta_{t}\sim\mathcal{N}(0,\sigma_{\omega ij})
\end{align*}
with $\sigma_{h}$ and $\sigma_{\omega ij}$ denoting the state-equation innovation variances.\footnote{In the empirical application, the likelihood turns out to be quite flat for $\sigma_{h}$, and we therefore impose the restriction $\sigma_{h}=0.2$. Evaluating various values for this parameter over a grid suggests this choice to be of minor importance.} Note that for the case of $\sigma_{h}$ and $\sigma_{\omega ij}$ equal to zero, we obtain homoscedastic errors.

Similar to \citet{crespo2017macroeconomic} and \citet{carriero2016measuring}, we interpret $h_t$ as the common driving force of the volatilities of all included series, and thus a measurement of uncertainty \citep[see also][]{jurado2015measuring}. Notice that $\text{Var}(\bm{\epsilon}_{t}) = \exp(h_t)\bm{L}\bm{L}' + \bm{\Omega}_t$, and the variance thus discriminates between idiosyncratic shocks and overall movements in international uncertainty. This feature can be linked to \citet{mumtaz2019evolving}, who rely on international, regional and country-specific factors. Our approach differs in the sense that we are only interested in international uncertainty $h_t$, and the remainder $\omega_{ij,t}$ captures both region and country-specific uncertainty.

The dynamic evolution of $\bm{y}_{it}$ is governed by a vector autoregressive (VAR) process with drifting coefficients \citep[similar to][]{doi:10.1111/rssa.12439} and features the common volatility of the factors in the mean:
\begin{equation}
\bm{y}_{it} = \bm{\alpha}_{it} + \sum_{p=1}^{P} \bm{A}_{ip,t} \bm{y}_{it-p} + \sum_{q=1}^{Q} \bm{B}_{iq,t} \bm{y}^{\ast}_{it-q} + \bm{\beta}_{it} h_t + \bm{\epsilon}_{it}.\label{eq:GVAR}
\end{equation}
We define the $k\times1$ intercept vector $\bm{\alpha}_{it}$ and $k\times k$ coefficient matrices $\bm{A}_{ip,t}~(p=1,\hdots,P)$. To establish dynamic interdependencies between economies in the spirit of \citet{Pesaran2004}, we construct a $k\times1$-vector $\bm{y}_{it}^{\ast} = \sum_{j = 1}^{N} w_{ij} \bm{y}_{jt}$. The $w_{ij}$ denote pre-specified weights (we let $w_{ii} = 0$, $w_{ij} \geq 0$ and $\sum_{j=1}^{N} w_{ij} = 1$ for $i,j=1,\hdots,N$) that capture the strength of the linkages. The process in Eq. (\ref{eq:GVAR}) is augmented by $Q$ lags of these non-domestic cross-sectional averages $\bm{y}_{it}^{\ast}$, with $k\times k$ coefficient matrices $\bm{B}_{iq,t}~(q=1,\hdots,Q)$. The vector $\bm{\beta}_{it}$ associated with the log of the factor volatility $h_t$ is of dimension $k\times1$.

Our setup allows for interpreting $\bm{\beta}_{it}$ as the contemporaneous impact of uncertainty $h_t$ on the endogenous variables of country $i$.\footnote{Note that one may in principle also include lags of $h_t$ in the mean of the process. Doing so, however, significantly complicates the sampling algorithm and we thus solely study its contemporaneous effects. Moreover, considering the state equation of $h_t$, its lag structure is implicitly featured in the VAR measurement equations subject to a set of parametric restrictions. Experiments including uncertainty as observed variable also featuring lags \citep[as in][]{doi:10.3982/ECTA6248,jurado2015measuring} do not affect the results from a qualitative viewpoint.} The structure set forth in Eqs. (\ref{eq:error}) and (\ref{eq:GVAR}) implies that shocks to $h_t$ affect both the first and second moments of the system based on common shocks captured by $\bm{f}_t$. We exploit this notion for calculating impulse response functions, relating to recursive identification schemes that order uncertainty indices first \citep[see, e.g.][]{doi:10.3982/ECTA6248}. In general, structural identification of uncertainty shocks is a challenging task due to various reasons, as suggested in \citet{ludvigson2015uncertainty}. Empirical evidence for the credibility of the identifying assumption employed in this paper in terms of economic interpretation is provided by \citet{carreiro2018endogenous}, who find little evidence for endogenous responses of macroeconomic uncertainty. In principle, the adopted setup would also allow to impose zero restrictions on the contemporaneous responses of lower-frequency real macroeconomic quantities. However, in the empirical application, we refrain from doing so and leave it up to the data whether these variables respond contemporaneously to uncertainty shocks.

Before proceeding, we recast the model in standard regression form,
\begin{align}
\bm{y}_{it} &= \bm{C}_{it} \bm{x}_{it} + \bm{\epsilon}_{it},\label{eq:GVARsimple}\\
\bm{x}_{it} &=(1,\bm{y}_{it-1}',\hdots,\bm{y}_{it-P}',\bm{y}^{\ast}{'}_{it-1},\hdots,\bm{y}^{\ast}{'}_{it-Q},h_t)'\nonumber\\
\bm{C}_{it} &= (\bm{\alpha}_{it},\bm{A}_{i1,t},\hdots,\bm{A}_{iP,t},\bm{B}_{i1,t},\hdots,\bm{B}_{iQ,t},\bm{\beta}_{it}).\nonumber
\end{align}
It is convenient to consider the $j$th equation of country $i$ in Eq. (\ref{eq:GVARsimple}) given by
\begin{equation*}
y_{ij,t} = \bm{C}_{ij,t}' \bm{x}_{it} + \epsilon_{ij,t}.
\end{equation*}
Here, we refer to the $j$th row of the matrix $\bm{C}_{it}$ by $\bm{C}_{ij,t}$, a vector of dimension $\tilde{K}\times1$ with $\tilde{K}=k(P+Q)+2$. The state vector is assumed to follow a random walk process
\begin{equation}
\bm{C}_{ij,t} = \bm{C}_{ij,t-1} + \bm{u}_t, \quad \bm{u}_t\sim\mathcal{N}(\bm{0},\bm{\Theta}_{ij}),\label{eq:stateEQ}
\end{equation}
with diagonal $\tilde{K}\times\tilde{K}$ variance-covariance matrix $\bm{\Theta}_{ij}=\text{diag}(\theta_{ij,1},\hdots,\theta_{ij,\tilde{K}})$. 

As for the stochastic volatility specification, if $\theta_{ij,l}$ equals zero in Eq. (\ref{eq:stateEQ}), the respective coefficient is constant over time. To test the restriction $\theta_{ij,l}=0$, we introduce the non-centered parameterization set forth by \citet{FRUHWIRTHSCHNATTER201085}, which allows to impose shrinkage priors on these innovation variances. In particular, this approach splits the model coefficients into a constant and a time-varying part, a feature we exploit for achieving shrinkage in the high-dimensional multivariate system.\footnote{For applications of this approach in a VAR context see \citet{bittosfs} and \citet{huber2019inducing}.}

Using a $\tilde{K}\times1$-vector containing the square root of the state innovation variances in Eq. (\ref{eq:stateEQ}) denoted $\sqrt{\bm{\Theta}_{ij}}=\text{diag}(\sqrt{\theta_{ij,1}},\hdots,\sqrt{\theta_{ij,\tilde{K}}})$, the reparameterized equation is
\begin{equation}
y_{ij,t} = \bm{C}_{ij,0}' \bm{x}_{it} + \tilde{\bm{C}}_{ij,t}' \sqrt{\bm{\Theta}_{ij}} \bm{x}_{it} + \epsilon_{ij,t}.\label{eq:GVAReqtrans}
\end{equation}
Let $\tilde{c}_{ijl,t}$ denote a typical element of $\tilde{\bm{C}}_{ij,t}$, then the transformation $c_{ijl,t} = c_{ijl,0} + \sqrt{\theta_{ij,l}} \tilde{c}_{ijl,t}$ yields the corresponding state equation
\begin{equation*}
\tilde{\bm{C}}_{ij,t} = \tilde{\bm{C}}_{ij,t-1} + \bm{v}_t, \quad \bm{v}_t\sim\mathcal{N}(\bm{0},\bm{I}_{\tilde{K}}),
\end{equation*}
with $\tilde{\bm{C}}_{ij,0} = \bm{0}_{\tilde{K}}$. This procedure moves the square root of the innovation variances to the states into Eq. (\ref{eq:GVAReqtrans}). The resulting state-space representation has the convenient property that the $\sqrt{\theta_{ij,l}}$ can conditionally be treated as standard regression coefficients.

Stochastically selecting which series should feature time-variation in their respective volatilities can be carried out using a similar transformation \citep[see][]{FRUHWIRTHSCHNATTER201085,KASTNER2014408}. Conditional on $\bm{L}\bm{f}_t$ and the full history of the VAR coefficients $\bm{C}_{it}$, we obtain a set of unrelated heteroscedastic error terms $\bm{\eta}_{t}$ by the diagonal structure of $\bm{\Omega}_t$. Here, $\eta_{ij,t}$ indicates the error term of the $j$th equation for country $i$. Squaring and taking logs and using $\omega_{ij,t} = \sqrt{\sigma_{\omega ij}} \tilde{\omega}_{ij,t}$ yields
\begin{align*}
&\tilde{\eta}_{ij,t} = \sqrt{\sigma_{\omega ij}} \tilde{\omega}_{ij,t} + \nu_{ij,t}, \quad \nu_{ij,t}\sim\ln\chi(1),\\
&\tilde{\omega}_{ij,t} = \tilde{\omega}_{ij,t-1} + w_{ij,t}, \quad w_{ij,t}\sim\mathcal{N}(0,1),
\end{align*}
again moving the square root of the innovation variances $\sqrt{\sigma_{\omega ij}}$ from the state to the measurement equation. The transformation allows to impose shrinkage priors on these coefficients, potentially pushing the model towards a homoscedastic specification if suggested by likelihood information.

\subsection{Prior distributions}
Bayesian methods are employed for estimation and inference. The panel structure of the data allows for constructing flexible shrinkage priors that are equipped to extract both cross-sectional information and shrink the model towards sparsity. Before proceeding we stack the coefficients using $\bm{c}_{i} = \text{vec}(\bm{C}_{i1,0}',\hdots,\bm{C}_{ik,0}')$. In similar fashion, we collect square roots of the innovation variances $\sqrt{\theta_{ij,l}}$ in
\begin{equation*}
\sqrt{\bm{\theta}_i} = (\sqrt{\theta_{i1,1}},\hdots,\sqrt{\theta_{i1,\tilde{K}}},\hdots,\sqrt{\theta_{ik,1}},\hdots,\sqrt{\theta_{ik,\tilde{K}}})',
\end{equation*} 
and index the $j$th elements by $c_{ij}$ and $\sqrt{\theta_{ij}}$, respectively, with $j=1,\hdots,k\tilde{K}$.

This article draws from the literature on the Bayesian treatment of panel data and global-local priors. In particular, we center the prior for all countries on a common mean that is estimated from the data, reflecting the notion that macroeconomic dynamics across economies are typically similar. The prior setup thus relates to the random coefficients and heterogeneity model \citep{verbeke1996linear,fruhwirth2004bayesian}, and restrictions often imposed in the context of panel VARs \citep[see][]{ecbwp:1507,KOOP2016115}. We propose hierarchical priors akin to the Normal-Gamma (NG) shrinkage prior of \citet{griffin2010} recently adopted in the VAR context by \citet{doi:10.1080/07350015.2016.1256217}. 

Since an analogous setup is applied for different parts of the parameter space, we rely on the generic indicator $\bullet$ for the indexes $\{c,\theta,\mu_c,\mu_\theta,\sigma,L\}$. For the constant part of the coefficients, we assume
\begin{equation}
c_{ij}|\mu_{cj},\tau_{cj}\sim\mathcal{N}(\mu_{cj},\tau_{cj}), \quad \tau_{cj}|\lambda_{c}\sim\mathcal{G}(a_\bullet,a_\bullet\lambda_{c}/2), \quad \lambda_{c}\sim\mathcal{G}(d_{\bullet0},d_{\bullet1}).\label{eq:NG1}
\end{equation}
Here, a key novelty is that we push all country-specific coefficients towards a common mean $\mu_{cj}$. The overall degree of shrinkage is determined by a global shrinkage parameter $\lambda_{c}$ serving as a general indicator of cross-country homogeneity. To provide flexibility for country-specific macroeconomic dynamics, we introduce local scaling parameters $\tau_{cj}$. In the presence of heavy shrinkage governed by $\lambda_{c}$, the $\tau_{cj}$ allow for flexibly selecting idiosyncrasies in coefficients across economies. This is an innovation compared to similar approaches \citep[see, e.g.][]{Malsiner-Walli2016,fischer2019regional} who solely rely on a set of Gamma priors on these variances, disregarding a common degree of overall shrinkage towards homogeneity.

Shrinkage on the innovation variances of the states in Eq. (\ref{eq:stateEQ}) is introduced in similar vein. We follow \citet{bittosfs} and stipulate a Gamma prior on these variances, which combined with a hierarchical prior relying again on Gamma distributions yields the setup they term the double Gamma prior. This is advantageous to the often employed inverse Gamma prior, because it does not artificially pull mass away from zero, a crucial feature when interest centers on stochastically shrinking the time-varying coefficients towards constancy. \citet{FRUHWIRTHSCHNATTER201085} show that this is equivalent to imposing a Gaussian prior on the square root of the state innovation variances,
\begin{equation*}
\sqrt{\theta_{ij}}|\mu_{\theta j},\tau_{\theta j}\sim\mathcal{N}(\mu_{\theta j},\tau_{\theta j}), \quad \tau_{\theta j}|\lambda_{\theta}\sim\mathcal{G}(a_{\bullet},a_{\bullet}\lambda_{\theta}/2), \quad \lambda_{\theta}\sim\mathcal{G}(d_{\bullet0},d_{\bullet1}).
\end{equation*}
As in the case of the constant coefficients of the model, we introduce a common mean $\mu_{\theta j}$ rather than pushing the variances towards zero a priori. This feature captures the notion that not only the constant coefficients across countries may be similar, but also the degree of time variation of the model parameters.

The first hierarchy of priors address the notion that the dynamic coefficients of the model might be similar over the cross-section. However, VARs with drifting coefficients are prone to overfitting issues. We deal with this problem and induce shrinkage in the coefficient matrices by imposing another NG prior to achieve regularization at the second level of the hierarchy. On the common mean $\mu_{sj}$ (for $s\in\{c,\theta\}$) we specify 
\begin{equation*}
\mu_{sj}|\tau_{\mu_s j}\sim\mathcal{N}(0,\tau_{\mu_s j}), \quad \tau_{\mu_s j}|\lambda_{\mu_s}\sim\mathcal{G}(a_{\bullet},a_{\bullet}\lambda_{\mu_s}/2), \quad \lambda_{\mu_s}\sim\mathcal{G}(d_{\bullet0},d_{\bullet1}).
\end{equation*}
This setup pushes its elements towards zero, where the overall level of shrinkage is governed by the global parameter $\lambda_{\mu_s}$. Similar to the first prior hierarchy, the prior allows for non-zero elements if suggested by the data via the local scalings $\tau_{\mu_s j}$.

For the stochastic volatility specification we rely on analogous priors. In particular, for the state innovation variances of the stochastic volatility processes for the $j$th variable of country $i$, we impose Gamma distributed priors, translating to Gaussian priors on the square root of these variances. The prior is given by
\begin{equation*}
\sqrt{\sigma_{\omega ij}}|\tau_{\sigma ij}\sim\mathcal{N}(0,\tau_{\sigma ij}), \quad \tau_{\sigma ij}|\lambda_\sigma\sim\mathcal{G}(a_\bullet,a_\bullet\lambda_\sigma/2), \quad \lambda_\sigma\sim\mathcal{G}(d_{\bullet0},d_{\bullet1}),
\end{equation*}
with the global shrinkage parameter $\lambda_\sigma$ pushing the model towards a homoscedastic specification. The local scalings $\tau_{\sigma ij}$ allow for non-zero state innovation variances. Intuitively, if $\tau_{\sigma ij}$ is small, we introduce substantial prior information and the parameter is pushed towards zero, ruling out time-variation in the respective volatility. For larger values of $\tau_{\sigma ij}$, the prior is less informative and allows for movements in the corresponding error variances.

It remains to specify prior distributions on the factor loadings in $\bm{L}$. Following \citet{kastner2019sparse}, we stack the elements in a vector $\bm{l}$ with typical element $l_j$ for $j=1,\hdots,R~(R=Kd)$ and impose
\begin{equation}
l_{j}|\tau_{Lj}\sim\mathcal{N}(0,\tau_{Lj}), \quad \tau_{Lj}|\lambda_{L}\sim\mathcal{G}(a_\bullet,a_\bullet\lambda_L/2), \quad \lambda_L\sim\mathcal{G}(d_{\bullet0},d_{\bullet1}).\label{eq:NG5}
\end{equation}

Until now we remained silent on the choices of hyperparameter values. In the empirical specification, and referring by $\bullet$ to the indexes $\{c,\theta,\mu_c,\mu_\theta,\sigma,L\}$, we follow the literature and set $d_{\bullet0}=d_{\bullet1}=0.01$ implying heavy shrinkage via the global parameter. Note that the hyperparameter $a_\bullet$ plays a crucial role regarding shrinkage properties. In fact, setting $a_\bullet=1$ would yield the Bayesian LASSO used in \citet{doi:10.1002/for.2276}. Given that the generic prior is applied to a range of different quantities of the model's parameter space, we integrate out this hyperparameter by imposing exponential priors $a_\bullet\sim\mathcal{E}(1)$.

Full conditional posterior distributions obtained from combining the likelihood function with the priors and the corresponding estimation algorithm are discussed in Appendices A and B. Fortunately, most distributions are of well-known form, allowing for a simple Markov chain Monte Carlo (MCMC) algorithm to obtain draws from the joint posterior using Gibbs sampling.

\section{Data and model specification}\label{sec:data}
Our dataset consists of monthly data for the period ranging from 1991:04 to 2018:07 for six economies: France (FRA), Germany (DEU), the United Kingdom (GBR), Italy (ITA), Japan (JPN), and the United States (USA). Consequently, the information set covers the G7 countries except Canada due to limitations of government bond yield data. The model features series on industrial production (IP, as a monthly indicator of economic activity), unemployment (UN), year-on-year consumer price inflation (PR), exports (EX) and equity prices (EQ). Industrial production, exports and equity prices enter the model in natural logarithms. To construct the cross-sectional weights for establishing links between economies, we rely on bilateral annual trade flows averaged over the sample period. Moreover, we obtain data on government bond yields at different maturities. The data are downloaded from the FRED database of the Federal Reserve Bank of St. Louis (\href{https://fred.stlouisfed.org/}{fred.stlouisfed.org}) and Quandl (\href{https://www.quandl.com/}{quandl.com}).\footnote{The period considered for this paper is dictated by the availability of the government bond yields at different maturities.}

A crucial determinant of business cycle fluctuations and the transmission of uncertainty shocks to the real sector of the economy are financial markets \citep{gilchrist2009credit,gilchrist2012credit,gilchrist2014nber,ALESSANDRI2018}. For a parsimonious representation of the full term structure of interest rates, we adopt a Nelson-Siegel type model \citep[see][]{10.2307/2352957,DIEBOLD2006337}. Government bond yield curves are estimated employing a factor model denoting yields by $r_{it}(\tau)$ at maturity $\tau$,
\begin{equation}
r_{it}(\tau) = \mathfrak{L}_{it} + \mathfrak{S}_{it} \left(\frac{1-\exp(-\lambda\tau)}{\lambda\tau}\right) + \mathfrak{C}_{it} \left(\frac{1-\exp(-\lambda\tau)}{\lambda\tau}-\exp(-\lambda\tau)\right).\label{eq:nelson-siegel}
\end{equation}
This setup allows the factors $\mathfrak{L}_{it}$, $\mathfrak{S}_{it}$ and $\mathfrak{C}_{it}$ to be interpreted as the level, (negative) slope and curvature of the yield curve, and may be estimated using ordinary least squares.\footnote{We adopt a two-stage procedure to reduce the computational burden. The factor loadings are determined by the parameter $\lambda=0.0609$ \citep[see][for details on this choice]{DIEBOLD2006337}.} Using an $m\times1$-vector of macroeconomic indicators $\bm{m}_{it}$, we exploit the yield curve fundamentals extracted in Eq. (\ref{eq:nelson-siegel}) to construct the $k\times1$ endogenous vector $\bm{y}_{it} = (\bm{m}_{it}',\mathfrak{L}_{it},\mathfrak{S}_{it},\mathfrak{C}_{it})'$ for $t=1,\hdots,T$ for country $i=1,\hdots,N$. In the discussion of the empirical results, $\mathfrak{L}_{it}$, $\mathfrak{S}_{it}$ and $\mathfrak{C}_{it}$ are labeled NSL, NSS and NSC, respectively. Thus, all dimensions of the involved vectors can be derived based on $k=8$ and $N=6$. 

To select the lag order of the model and the number of latent factors that drive the full system covariance matrix, we rely on the deviance information criterion (DIC). We estimate the model over a grid of lag and latent factor combinations, and choose the specification minimizing the DIC. This procedure selects a model with $P=Q=2$ lags and $d=4$ factors. 

For the empirical application, we adopt the general prior setup put forward in Section \ref{sec:econometrics}. In particular, to reduce influence of the prior setup on the estimated impact of uncertainty, we use a rather diffuse prior on the constant part of these coefficients with prior variance equal to ten. The square roots of the state innovation variances of the impact vector are tightly centered on zero a priori. The latter choice mutes differences in impact reactions over time, but improves the stability of the model.

\section{Empirical results}\label{sec:results}
\subsection{The measure of uncertainty}
\begin{figure}[t]
\begin{center}
\includegraphics[width=\textwidth]{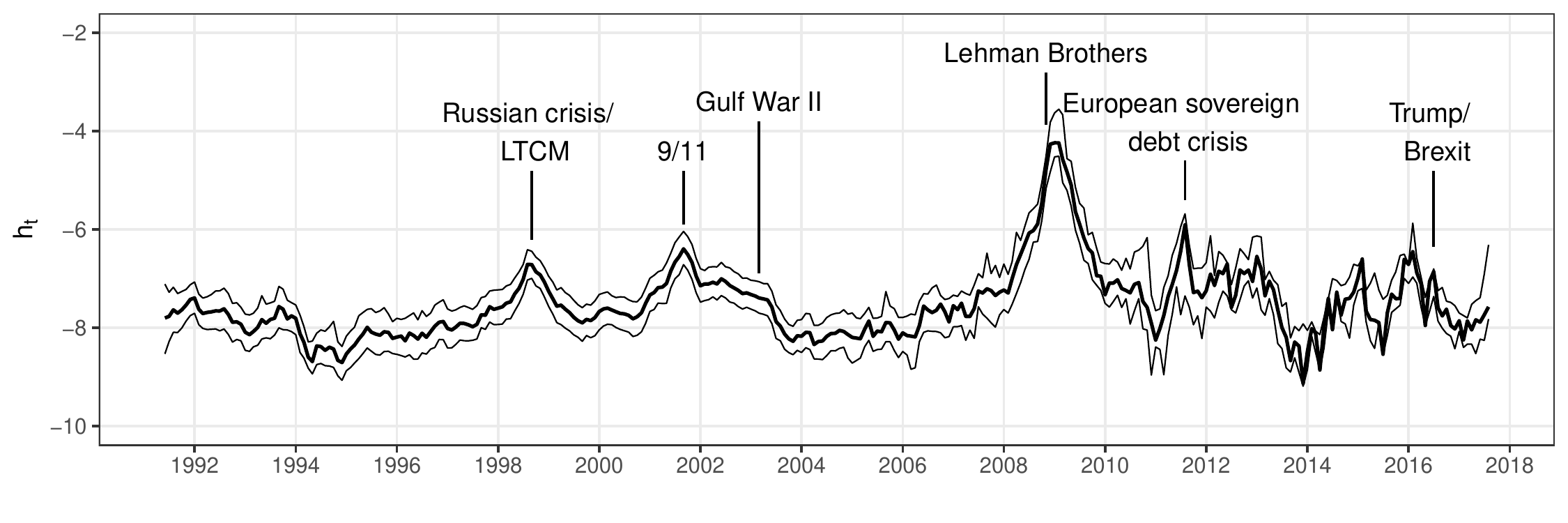}
\end{center}
\caption{Measurement of uncertainty depicting the log-volatility $h_t$ of the factors.}\label{fig:unc1}\vspace*{-0.3cm}
\caption*{\footnotesize\textit{Note}: The thick black line depicts the posterior median, alongside the $16$th and $84$th posterior percentiles (thin lines). LTCM is the collapse of Long-Term Capital Management, 9/11 indicates the terror attacks of September 11, 2001.}
\end{figure}

Figure \ref{fig:unc1} displays the estimated measure of uncertainty: the log-volatility $h_t$ driving the common shocks to the system. The most striking episode of international uncertainty occurs during the global financial crisis and subsequent Great Recession. Several less pronounced episodes of similarly elevated levels are worth noting. Chronologically, uncertainty rises in the first half of 1997, related to the Asian financial crisis. A spike in late 1998 reflects the Russian financial crisis and the subsequent collapse of the U.S. hedgefund Long-term Capital Management. Afterwards, a brief period of lower uncertainty is observable, coming to an end with the burst of the Dot-com bubble and the 9/11 terror attacks in late 2001. Sustained elevated levels, albeit declining, are observable until the end of 2003, a period encompassing the second Gulf War. The period between 2004 and the bancruptcy of the U.S. investment bank Lehman Brothers features lower levels of international uncertainty. 

Surging international volatilities are detected by the model starting in late 2007, capturing the onset of the crisis in the U.S. subprime mortgage market and first signs of disturbances on credit markets. After a decline of common volatilities to pre-crisis levels around 2010, the second highest peak of $h_t$ occurs in 2011, related to events during the European sovereign debt crisis. This period of elevated uncertainty is sustained until late 2013. The most recent episode of high uncertainty emerges in early 2016, indicating peaks related to the Brexit referendum and the election of Donald Trump as President of the United States in late 2016.

\begin{figure}[!htbp]
\begin{center}
\includegraphics[width=\textwidth]{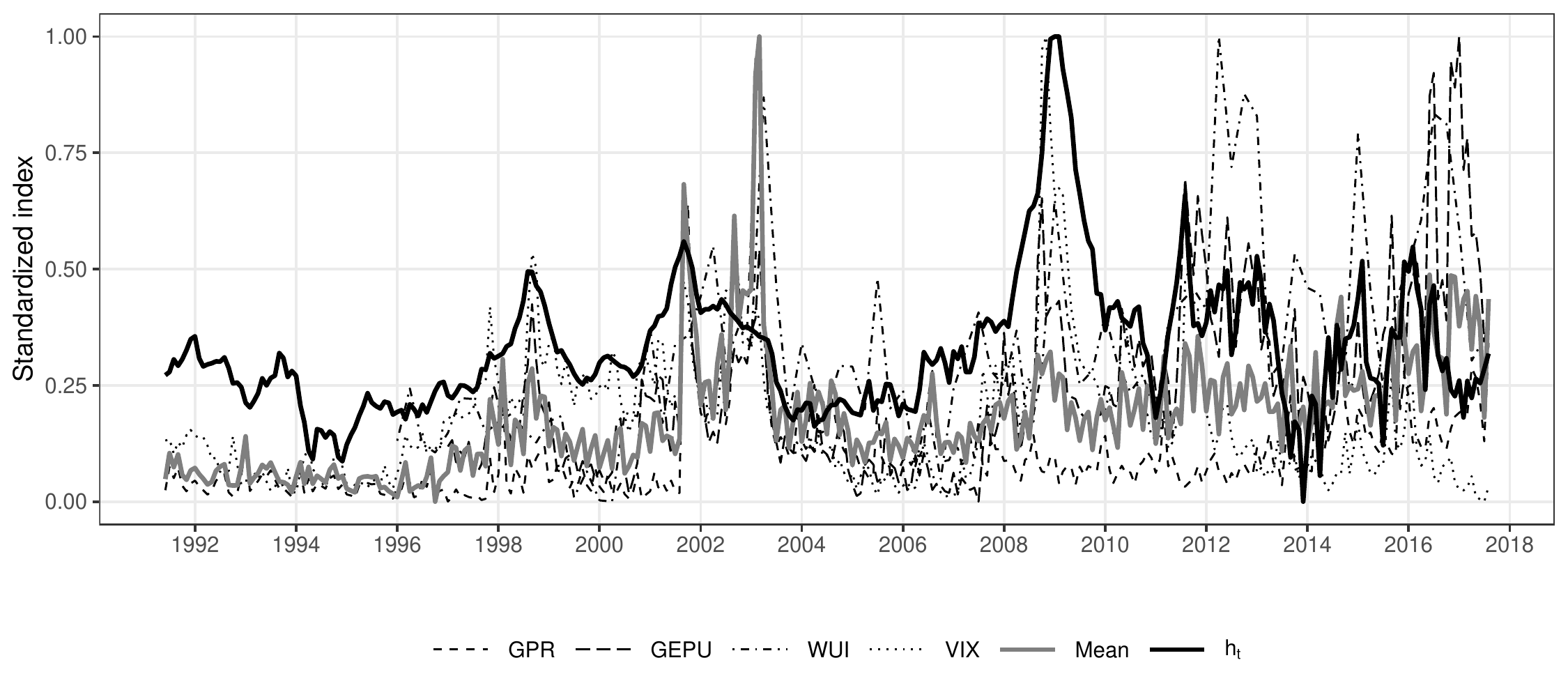}
\end{center}
\caption{Comparison of standardized uncertainty measures over time.}\label{fig:unc2}\vspace*{-0.3cm}
\caption*{\footnotesize\textit{Note}: Measures are standardized to lie in the unit interval. The thick black line depicts the posterior median of $h_t$. Uncertainty measures:  Geopolitical risk (GPR), global policy uncertainty (GEPU), world uncertainty index (WUI), CBOE volatility index (VIX).}
\end{figure}

Following this brief discussion of the measure in light of uncertainty-related events, we compare our findings to commonly adopted proxies for uncertainty. We consider the geopolitical risk (GPR) index described in \citet{caldara2018measuring}, the global policy uncertainty (GEPU) index and the world uncertainty index (WUI) constructed as described in \citet{doi:10.1093/qje/qjw024}, and complement them with the Chicago Board Options Exchange (CBOE) volatility index (VIX).\footnote{The indices are available for download at \href{https://www2.bc.edu/matteo-iacoviello/gpr.htm}{www2.bc.edu/matteo-iacoviello/gpr.htm} (GPR), \href{https://www.policyuncertainty.com/}{policyuncertainty.com} (GEPU and WUI) and the FRED database of the Federal Reserve Bank of St. Louis (VIX).} Moreover, we take the arithmetic average for all benchmark indices and label the resulting series ``Mean'' in corresponding visualizations. To make the scales of the uncertainty measurements comparable, we standardize all measures to lie in the unit interval. 

The resulting series are depicted in Fig. \ref{fig:unc2}. A few points are worth noting. First, $h_t$ provides a smoother estimate of uncertainty. However, most peaks apparent in the benchmark uncertainty measures are traced accurately. Differences occur mainly in the magnitude of the implied level of uncertainty. For instance, ``Mean'' peaks in 2003, with most benchmark measures showing substantial uncertainty around the outbreak of the second Gulf War. The endogenous measure of uncertainty traces this peak, but at a comparatively lower level. The Great Recession peak in late 2008 on the other hand, exhibiting a spike in the VIX and most other measures apart from GPR, is the highest level of uncertainty detected by $h_t$. Maximum values of WUI are associated with elevated levels in $h_t$, and also the peaks of GPR and GEPU coincide with upward movements in $h_t$. It is worth mentioning that our uncertainty measurement compares well to similar approaches dealing with the endogenous measurement of uncertainty \citep[see][]{crespo2017macroeconomic,carriero2018assessing}.

\begin{figure}[t]
\begin{center}
\includegraphics[width=\textwidth]{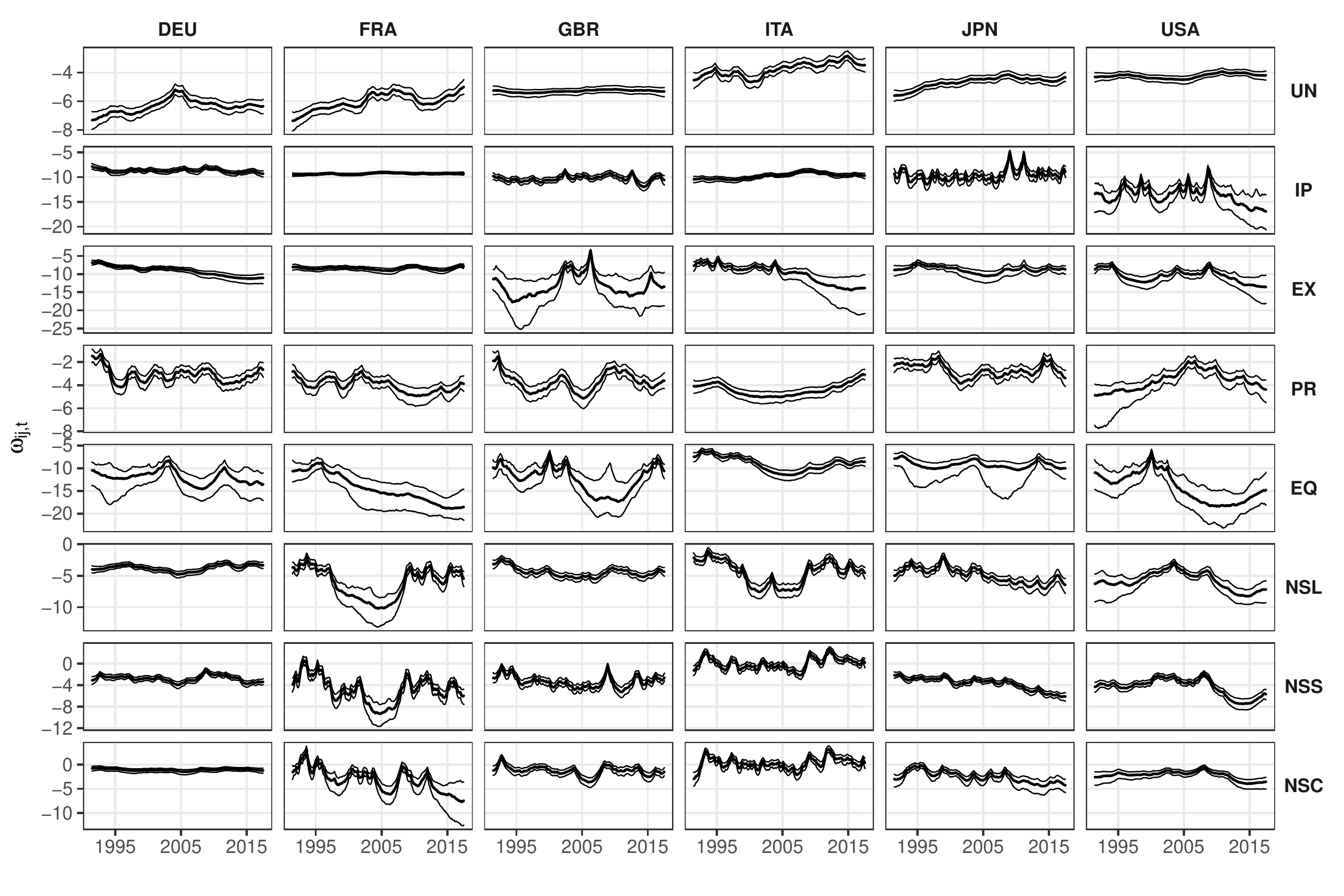}
\end{center}
\caption{Series-specific log-volatilities $\omega_{ij,t}$.}\label{fig:S}\vspace*{-0.3cm}
\caption*{\footnotesize\textit{Note}: The thick black line depicts the posterior median, alongside the $16$th and $84$th posterior percentiles (thin lines).}
\end{figure}

The discussion is complemented by the findings for idiosyncratic volatility series. Recall that the prior setup imposes shrinkage on the idiosyncratic residual variances towards constancy. As evidenced by \autoref{fig:S}, the likelihood strongly suggests the necessity of a stochastic volatility specification. Individual series feature pronounced heterogeneities both in terms of the magnitude and the timing of peaks. This provides evidence that the approach employed for measuring common uncertainty in this paper discriminates well between country-specific events and international uncertainty-related events of significance.

Largest differences in the magnitude of the volatilities are visible for unemployment, with Germany and France exhibiting lower residual variances. However, both feature substantial higher-volatility periods in the years surrounding 2005. While $\omega_{ij,t}$ for industrial production is rather homogenous for the continental European countries, the series of the remaining economies exhibit heterogeneities both in terms of magnitude and time-variation. The same is true, even though to a slightly lesser degree, in the case of export volatilities. Moreover, pronounced time-variation is clearly featured in the respective series relating to country-specific inflation dynamics, and equity prices. Volatilities associated with the factors capturing yield curve dynamics show marked similarities across countries, reflecting international commonalities in equity markets.

\subsection{Dynamic responses to uncertainty shocks}
The non-centered parameterization of the state-space model in principle allows to investigate both shrinkage on the common mean of the time-invariant part of the model coefficients, and the corresponding state innovation variances. Moreover, the degree of shrinkage towards homogeneity over the cross-section can be assessed. We find that for the most part, the parameter space is pushed toward both cross-sectional homogeneity and time-invariance of the coefficients. This appears sensible, given the size of the system and the dicussion in \citet{feldkircher2017sophisticated} regarding time-varying coefficients often capturing omitted variable bias in smaller VARs. For the sake of brevity, we refrain from presenting detailed inference on the shrinkage parameters governing homogeneity over time and the cross-section and provide further information in Appendix C.

\begin{figure}[!ht]
\begin{center}
\includegraphics[width=\textwidth]{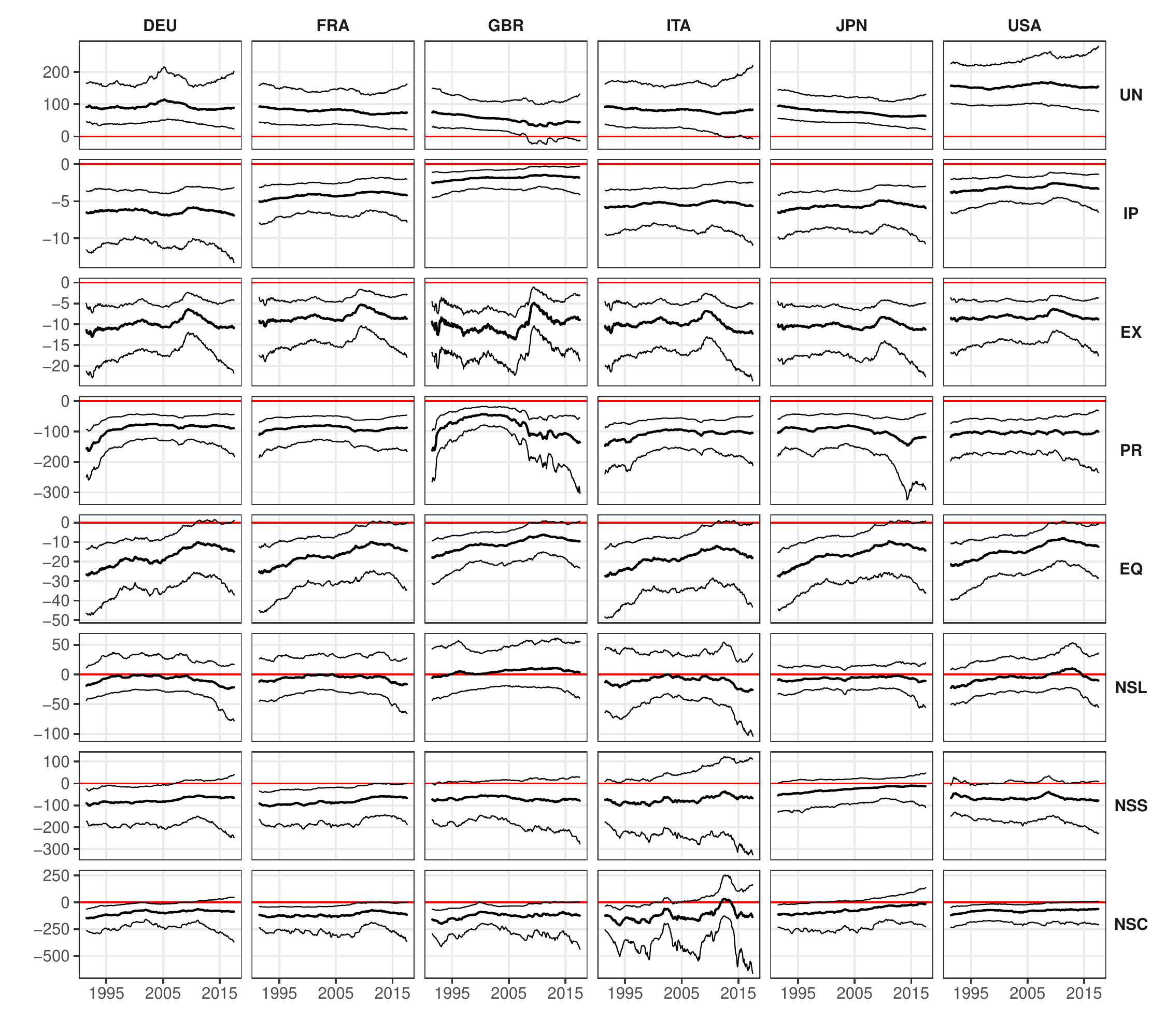}
\end{center}
\caption{Cumulative impulse response functions to an international uncertainty shock.}\label{fig:p_irfcum}\vspace*{-0.3cm}
\caption*{\footnotesize\textit{Note}: The thick black line depicts the posterior median, alongside the $16$th and $84$th posterior percentiles (thin lines). The red line marks zero.}
\end{figure}

Turning to structural inference, Fig. \ref{fig:p_alltime} displays an overall summary of the dynamic responses for the periods between January 1992 and July 2017 on a biannual frequency, and reports the posterior median of the impulse response functions to the uncertainty shock. Colors refer to the respective period (red indicates early parts of the sample, blue marks later periods). Figure \ref{fig:p_irfcum} depicts cumulative responses at the five year horizon. Units are scaled as percentages for industrial production, exports and equity prices, while consumer price inflation, unemployment and the Nelson-Siegel factors for level, slope and curvature are in basis points (BPs).\footnote{To preserve space, we refrain from presenting numerical results. Detailed tables are available upon request.}

In general, our results corroborate empirical findings from previous contributions, and both directions and magnitudes of the responses are similar. One notable result concerning the timing of the responses is that most react strongly on impact of the shock. We find significant increases of unemployment in all countries, while industrial production, exports, inflation and equity prices decrease. Timing and shape of the impulse responses for Nelson-Siegel level, slope and curvature factors indicate a flattening of the yield curve associated with overall decreases in interest rates at most maturities.

\begin{figure}[!ht]
\begin{center}
\includegraphics[width=\textwidth]{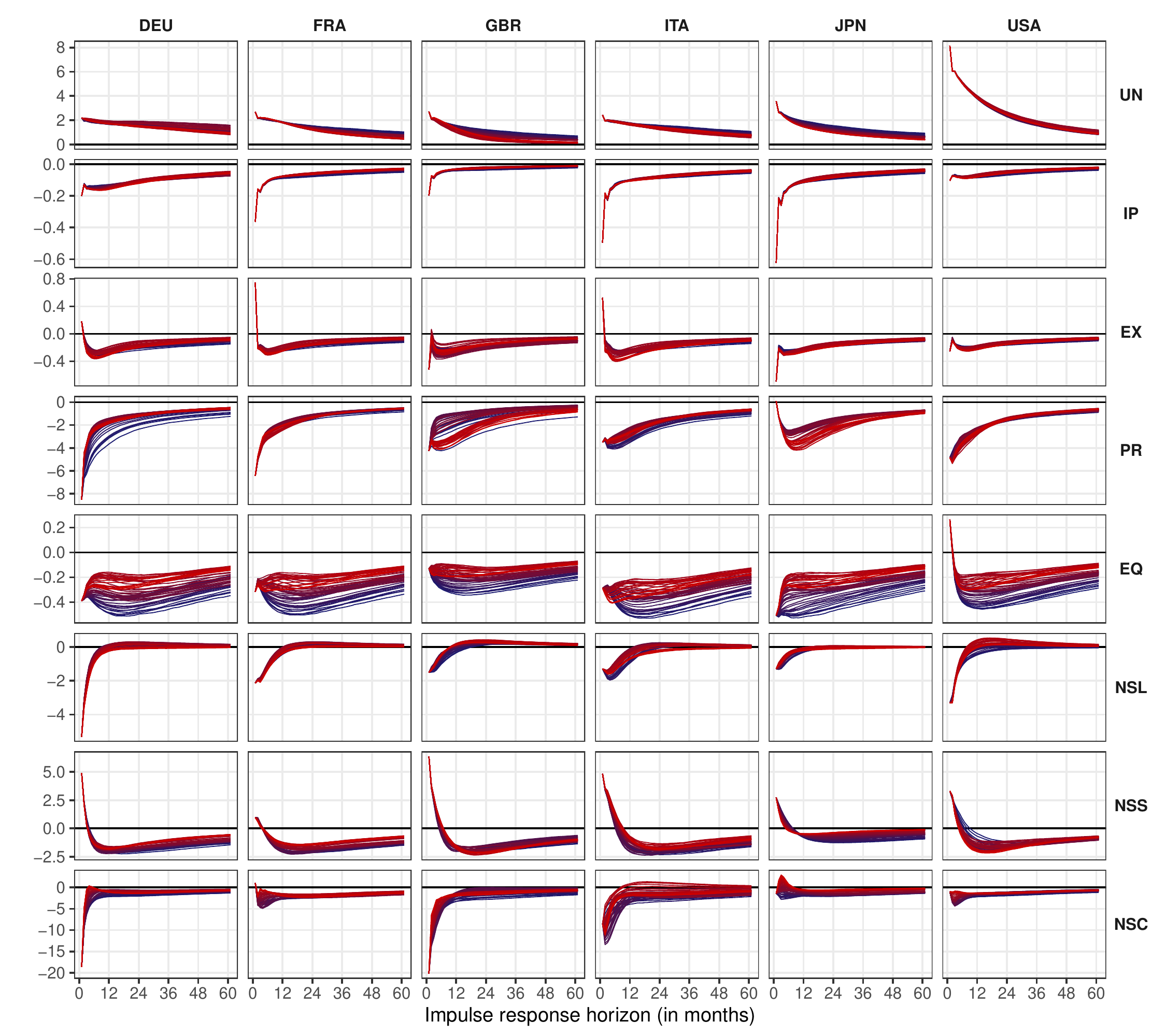}
\end{center}
\caption{Impulse responses for selected periods to an international uncertainty shock.}\label{fig:p_alltime}\vspace*{-0.3cm}
\caption*{\footnotesize\textit{Note}: Posterior median of the impulse response functions over time, with the shading referring to the respective period: \textcolor[HTML]{191970}{\textbf{------}} 1992:01 to \textcolor[HTML]{CD0000}{\textbf{------}} 2017:07 on biannual frequency. The black line marks zero.}
\end{figure}

\textit{Unemployment.}
For unemployment we detect significant peaks on impact, ranging from two BPs in the case of Germany, France, the United Kingdom and Italy, while Japan exhibits slightly larger magnitudes. The largest responses result in the United States, with increases up to eight BPs. The estimated effects are rather persistent, with significant positive reactions over the impulse response horizon of five years. Figure \ref{fig:p_alltime} suggests only a minor degree of time-variation, with the impacts leveling out slightly quicker in later parts of the sample.

Closer inspection of the cumulative effects in the first row of Fig. \ref{fig:p_irfcum} yields some interesting insights. Systematic decreases are visible for France, Italy, the United Kingdom and Japan. This notion is most pronounced for the United Kingdom. Different behavior occurs in Germany and the United States, with substantially larger cumulative responses. Estimates for the United States gradually amplify before the global financial crisis.

\textit{Industrial production.}
Industrial production shows the largest declines in Italy and Japan, with significant negative peak responses on impact. The remaining countries exhibit rather homogeneous responses of approximately $0.2$ percent, while the United States show the smallest effects, with a decline of $0.1$ percent on impact. Note that for some countries posterior credible sets of the peak responses include zero. Time variation appears limited, however, as reported in the second row Fig. \ref{fig:p_irfcum}, subtle changes in the persistence of the estimated uncertainty shocks translate to time-varying patterns in terms of cumulative responses.

Cumulative effects gradually decrease in the first years of the sample period, in line with \citet{mumtaz2017changing}. This trend disappears just prior to the global financial crisis. Interestingly, in a brief period after the Great Recession, uncertainty shocks appear to play a less important role for industrial production, a notion that reverts later in the sample.

\textit{Exports.}
Exports indicate insignificant contemporaneous effects close to zero for Germany and the United Kingdom, with a significant peak decline around two quarters after impact. France and Italy exhibit positive impacts, but the responses quickly turn negative. Substantial decreases are indicated for the United States and Japan. The magnitude of the impacts is in line with findings by \citet{crespo2017macroeconomic}; they are also interesting when interpreted in light of \citet{mumtaz2017common}, who suggest that international trade plays a crucial role in the transmission of uncertainty shocks via falling foreign demand and feedback volatility effects for the respective domestic economy.

No clear time-variation patterns emerge in Fig. \ref{fig:p_alltime}. Here, we again resort to Fig. \ref{fig:p_irfcum}, with the third row providing evidence of substantial differences in the cumulative responses over time. Prior to the global financial crisis, estimates fluctuate around $-10$ percent across countries. The largest fluctuations in the cumulative responses are observable for the United Kingdom. Analogous to the results for unemployment and industrial production, the consequences of uncertainty shocks on exports in the aftermath of the Great Recession are muted in comparison to previous periods. From 2015 onwards, effects are again similar to earlier in the sample, with minor differences over the cross-section.

\textit{Consumer price inflation.}
\citet{10.1257/aer.20121236} identify two contradicting channels how uncertainty affects consumer prices: The so-called aggregate demand channel, characterized by reducing the consumption of households and thereby leading to an overall decrease in prices; and the upward-pricing bias channel, which yields increases in inflation based on profit-maximizing firms. In our case, the former appears to dominate the latter, with significant decreases of inflation on impact for most economies in row four of Fig. \ref{fig:p_irfcum}. The estimated peak effects indicate constancy in magnitudes ranging from $-8.5$ BPs in Germany on impact, to a mere $-2.5$ BPs for the case of Japan after two quarters (with insignificant impact responses). The impulses for Germany, France, Italy and the United States exhibit only a small degree of persistence, with responses quickly leveling out. In terms of time-variation, the effects of uncertainty shocks on prices appear slightly more persistent early in the sample. The shape of inflation responses in the United Kingdom is similar to the other countries before 2005, however, impulse response functions turn hump-shaped in later periods, comparable to those of Japan.

Further inspection of the estimates in light of Fig. \ref{fig:p_irfcum} reveals substantial heterogeneities. First, we observe differences in posterior uncertainty over the sample period. Less precisely estimated cumulative effects mainly occur in the context of short-term interest rates hitting zero-lower bound for most economies. Inflated credible sets and differences in the posterior median moreover occur early in the sample. Second, responses at the end of the sample period in July 2017 feature little cross-sectional heterogeneity. Finally, idiosyncratic movements for the United Kingdom are worth mentioning. After large negative effects early in the sample, the consequences of uncertainty shocks on inflation declined substantially until late 2007. After the Great Recession, substantially larger effects are detected.

\textit{Equity prices.}
Equity prices, displayed in the fifth row of Fig. \ref{fig:p_irfcum}, prominently feature time-variation in the dynamic responses for all countries. The responses in the United Kingdom are less pronounced than in the other countries, peaking after roughly one and a half years at about $-0.3$ percent. For the United States, we observe an insignificant impact response quickly turning negative, with peak effects after one year between $-0.25$ and $-0.5$ percent, depending on the respective period. The remaining economies exhibit rather similar responses, with Japan indicating the largest impact responses.

The cumulative responses for equity prices in the beginning of the sample period show large and significant homogeneous declines over the cross-section of economies, except for the United Kingdom where the posterior median is substantially smaller. The cumulative responses for equity prices move towards zero across all economies rather homogeneously, declining in absolute values by approximately $5$ percentage points. At the end of the sample, the cumulative responses decline even further, with the $68$ percent posterior credible set covering zero in some economies. A clear empirical regularity is that the impact of uncertainty shocks on equity prices declines over time, in line with findings provided in \citet{mumtaz2017changing}. However, note that this downward trend is not linear for the whole sample period, with subtle changes especially in periods associated with economic crisis and higher international uncertainty.

\textit{Nelson-Siegel factors.}
Impulse responses for the Nelson-Siegel factors are displayed in the last three rows of Fig. \ref{fig:p_alltime}.\footnote{For interpretational clarity, recapture Eq. (\ref{eq:nelson-siegel}), where the loading on $\mathfrak{L}_{it}$ is a constant for all $\tau$; hence it affects all maturities equally, and is interpreted as the long-term level of the yield curve. The loading associated with $\mathfrak{S}_{it}$ decreases rapidly in $\tau$, and is thus closely related to the negative slope of the yield curve and term spreads \citep[for details, see][]{DIEBOLD2006337}. Consequently, an \textit{increase} in $\mathfrak{S}_{it}$ implies a \textit{decrease} in term spreads, and thus a flattening of the yield curve. The loading of $\mathfrak{C}_{it}$ is hump-shaped, and loads most strongly on the middle segment of the yield curve that affects its curvature.} The dynamic evolution of the level factor exhibits substantial heterogeneity across countries, but appears comparatively constant over time with slight differences in the curvature of the responses. In particular, we find the largest and significant decreases on impact, coinciding with the peak response, for Germany. In general, the credible sets associated with the impulse responses of the level factor are rather large, and cover zero in most economies. The effects peter out quickly, with impulse responses returning to zero after about two quarters. Observed heterogeneity over the cross-section may originate from international capital flows toward safer assets in uncertain times \citep[see, e.g.][]{caballero2017safe}. Figure \ref{fig:p_irfcum} indicates that the posterior distributions of the cumulative effects for the level factor cover zero for all economies over the sample period considered, featuring detectable yet insignificant time-varying dynamics in the responses.

The slope factor detects significant positive reactions peaking instantaneously in Germany, the United Kingdom, and Japan. The effects for the remaining countries on impact are estimated less precisely, however, the posterior centers on positive values for all countries ranging from one to five BPs. An increase in the slope factors translates to a decrease in term spreads and a flattening of the yield curve, a phenomenon that has been linked to the emergence of recessions. This effect reverses in subsequent months, turning significantly negative between one and one and a half years after impact across countries. Given the close empirical relationship between the slope factor and central bank policy \citep[see][]{diebold2006macroeconomy}, we conjecture that this pattern captures a delayed response of central banks, lowering policy rates to counteract detrimental economic effects of uncertainty shocks.

Assessing cumulative effects, we find that estimates are statistically significant early in the sample for Germany and France. The model captures large but insignificant effects for the remaining economies except for Japan, which is unsurprising considering the country's recent monetary history. In general, the impact of uncertainty shocks on the slope factor appears to decrease over time, evidenced by subtle trends visible for most countries except the United Kingdom and the United States. Linking this finding to less responsive equity markets, such dynamics may be explained based on the theoretical model in \citet{mumtaz2017changing}, who suggest differences in the conduct of central banking to be related to decreases in the effects of uncertainty shocks.

The results associated with the curvature factor signal decreases for most countries. Again, we observe pronounced heterogeneity over the cross-section, but also over time. The responses peak on impact for Germany and the United Kingdom. France, the United States and Japan show only small consequences of uncertainty shocks for middle-term maturities. In terms of cumulative responses, we find systematic declines in the magnitude of the effects associated with inflated posterior uncertainty for Japan, dynamics that are also visible in the case of Germany, France, the United Kingdom and the United States. Minor differences occur for selected periods after the Great Recession. Italy presents a special case, with distinct periods featuring substantial differences in the cumulative responses.

\section{Concluding remarks}\label{sec:conclusions}
The obtained measure of uncertainty is comparable to established proxies, and the factor stochastic volatility structure discriminates well between events confined to individual economies and overall macroeconomic uncertainty. Uncertainty shocks cause downward pressure on inflation, increase unemployment levels, decrease industrial production and depress equity prices, with differences in timing and magnitude of the effects across economies. The term structure of interest rates generally exhibits decreases in levels at all maturities, with an accompanying overall flattening of the yield curve. The consequences of uncertainty shocks appear to decline gradually for some macroeconomic and financial quantities, while other variables show little variation over time. We find limited evidence for abrupt changes in the transmission channels of uncertainty shocks.

\newpage
\small{\setstretch{0.85}
\addcontentsline{toc}{section}{References}
\bibliographystyle{custom.bst}
\bibliography{lit}}

\newpage
\begin{appendices}\crefalias{section}{appsec}
\setcounter{equation}{0}
\setcounter{figure}{0}
\renewcommand\theequation{A.\arabic{equation}}
\section*{Appendix A: Posterior distributions and algorithm}\label{app:posteriorsalgorithm}
Conditional on $\{\bm{f}_t\}_{t=1}^{T}$ and the loadings $\bm{L}$, the full system of equations reduces to $K$ unrelated regression models with heteroscedastic errors. This allows for estimation of the system equation-by-equation, greatly reducing the computational burden. To see this, define $\bm{\tilde{y}}_t=\bm{y}_t-\bm{L}\bm{f}_t$ and refer to the $j$th variable of country $i$ by $\tilde{y}_{ij,t}$,
\begin{equation*}
\tilde{y}_{ij,t} = \bm{C}_{ij,0}' \bm{x}_{it} + \tilde{\bm{C}}_{ij,t}' \sqrt{\bm{\Theta}_{ij}} \bm{x}_{it} + \eta_{ij,t}.
\end{equation*}
Moreover, conditional on $\{\tilde{\bm{C}}_{ij,t}\}_{t=1}^{T}$, the innovation variances in $\sqrt{\bm{\Theta}_{ij}}$ can be treated as standard regression coefficients. We define the vector $\bm{d}_{ij}=(\bm{C}_{ij,0}',\sqrt{\theta_{ij,1}},\hdots,\sqrt{\theta_{ij,\tilde{K}}})'$. Let $\bullet$ refer to conditioning on all the other parameters, latent states of the model, and the data; then the posterior distribution of $\bm{d}_{ij}$ is a multivariate Gaussian,
\begin{equation}
\bm{d}_{ij}|\bullet\sim\mathcal{N}(\bm{\tilde{\mu}}_{ij},\bm{\tilde{V}}_{ij}).\label{eq:posteriorVARcoef}
\end{equation}
The posterior moments are $\bm{\tilde{V}}_{ij} = (\bm{\tilde{X}}_{ij}'\bm{\tilde{X}}_{ij}+\bm{V}^{-1})^{-1}$ and $\bm{\tilde{\mu}}_{ij}=\bm{\tilde{V}}_{ij}(\bm{\tilde{X}}_{ij}'\bm{\tilde{Y}}_{ij} + \bm{V}^{-1}\bm{\mu})$, with prior moments $\bm{\mu}=(\mu_{c1},\hdots,\mu_{c\tilde{K}},\mu_{\theta1},\hdots,\mu_{\theta \tilde{K}})'$ and $\bm{V}=\text{diag}(\tau_{c1},\hdots,\tau_{c\tilde{K}},\tau_{\theta1},\hdots,\tau_{\theta \tilde{K}})$. The matrix $\bm{\tilde{X}}_{ij}$ is of dimension $T\times2\tilde{K}$, with the $t$th row given by $[\bm{x}_{it}',\tilde{\bm{C}}_{ij,t}'\odot\bm{x}_{it}']\exp(-\omega_{ij,t}/2)$, $\bm{\tilde{Y}}_{ij}$ is $T\times1$ with $t$th element $\tilde{y}_{ij,t}\exp(-\omega_{ij,t}/2)$. This normalization enables to draw the coefficients from standard posterior quantities for the parameters of homoscedastic linear regression models.

Given draws for the country-specific constant part of the model parameters and the state innovation variances, it is straightforward to obtain the conditional posterior distributions for the prior moments collected in $\bm{\mu}$ and $\bm{V}$. Since the results apply to the coefficients in $\bm{c}_i$ and $\sqrt{\bm{\theta}}_i$, we again use an indicator $s\in\{c,\theta\}$ and obtain
\begin{equation*}
\tau_{sj}|\bullet\sim\mathcal{GIG}\left(a_s-N/2,\sum_{i=1}^{N}(c_{ij}-\mu_{sj})^2,a_s\lambda_s\right), \quad \lambda_s|\bullet\sim\mathcal{G}\left(d_{s0}+k\tilde{K}a_s,d_{s1}+\frac{a_s}{2}\sum_{j=1}^{k\tilde{K}}\tau_{sj}\right),
\end{equation*}
with local scalings $\tau_{sj}$ following a generalized inverse Gaussian distribution and the global shrinkage parameter a Gamma distribution. Conditional on $\{c_{ij}\}_{i=1}^{N}$, standard methods yield a Gaussian posterior
\begin{equation*}
\mu_{sj}\sim\mathcal{N}(\tilde{\mu}_{sj},\tilde{V}_{sj}),
\end{equation*}
with $\tilde{V}_{sj}=(N\tau_{sj}^{-1}+\tau_{\mu_s j}^{-1})^{-1}$ and $\tilde{\mu}_{sj}=\tilde{V}_{sj}(\sum_{i=1}^N c_{ij}\tau_{sj}^{-1})$. For the prior variance of the common mean, $\tau_{\mu_s j}$ it is straightforward to obtain
\begin{equation*}
\tau_{\mu_sj}|\bullet\sim\mathcal{GIG}\left(a_{\mu_s}-1/2,\mu_{sj}^2,a_{\mu_s}\lambda_{\mu_s}\right), \quad \lambda_{\mu_s}|\bullet\sim\mathcal{G}\left(d_{\mu_s0}+k\tilde{K}a_{\mu_s},d_{\mu_s1}+\frac{a_{\mu_s}}{2}\sum_{j=1}^{k\tilde{K}}\tau_{\mu_sj}\right).
\end{equation*}

To obtain draws from the posterior distribution of $\sigma_{\omega ij}$ we rely on the methods discussed in \citet{KASTNER2014408}. Conditional on a realization of $\sigma_{\omega ij}$, we obtain
\begin{equation*}
\tau_{\sigma ij}|\bullet\sim\mathcal{GIG}(a_{\sigma}-1/2,\sigma_{\omega ij},a_\sigma\lambda_\sigma), \quad \lambda_{\sigma}\sim\mathcal{G}\left(d_{\sigma0}+Ka_{\sigma},d_{\sigma1}+\frac{a_\sigma}{2}\sum_{i=1}^{k}\sum_{j=1}^{N}\tau_{\sigma ij}\right).
\end{equation*}

Note that Eq. (\ref{eq:error}) conditional on the other parameters of the model is a simple linear regression model with conditionally homoscedastic errors and standard formulae apply. The NG prior employed for the factor loadings translates to the following posteriors for the corresponding global and local shrinkage parameters:
\begin{equation*}
\tau_{Lj}|\bullet\sim\mathcal{GIG}(a_{L}-1/2,l_j^2,a_L\lambda_L), \quad \lambda_{L}\sim\mathcal{G}\left(d_{L0}+Ra_{L},d_{L1}+\frac{a_L}{2}\sum_{j=1}^{R}\tau_{Lj}\right).
\end{equation*}

We proceed with the posterior distribution for the hyperparameters of the prior on the local scalings $a_\bullet$. For the sake of brevity we refrain from presenting all respective indices and refer to the various possible index combinations using $\bullet$. Combining likelihood and prior, the conditional posterior for this parameter has no well-known form and we rely on a Metropolis-Hastings step for simulation. Given the support of $a_\bullet$, we propose candidate draws $a_\bullet^{\ast}$ from $\mathcal{N}(\ln(a_\bullet),\kappa_\bullet)$, with $\kappa_\bullet$ denoting a tuning parameter that is updated during half of the burn-in period to achieve an acceptance rate between $0.15$ and $0.35$. Acceptance probabilities are given by
\begin{equation*}
\min\left[1,\frac{p(a_\bullet^{\ast})p(a_\bullet^{\ast}|\tau_{\bullet})a_\bullet^{\ast}}{p(a_\bullet)p(a_\bullet|\tau_{\bullet})a_\bullet}\right].
\end{equation*}

\renewcommand\theequation{B.\arabic{equation}}
\section*{Appendix B: MCMC algorithm}\label{app:mcmc}
Employing the posterior distributions presented in Appendix A, the full MCMC algorithm cycles through the following steps:
\begin{enumerate}
  \item The constant part of the coefficients and the process variances of the coefficients are simulated equation-by-equation using Eq. (\ref{eq:posteriorVARcoef}).
  \item For $\{\tilde{\bm{C}}_{ij,t}\}_{t=1}^{T}$, we rely on a forward filtering backward sampling algorithm \citep[see][]{doi:10.1093/biomet/81.3.541,doi:10.1111/j.1467-9892.1994.tb00184.x}.
  \item Conditional on the country-specific coefficients, it is straightforward to obtain a draw for the common mean $\bm{\mu}$ and the associated global and local shrinkage parameters in $\bm{V}$. Subsequently, given a simulated value for the common mean, we draw the global and local shrinkage parameters $\tau_{\mu_sj}$ and $\lambda_s$.
  \item Simulation of $\{\omega_{ij,t}\}_{t=1}^{T}$ is carried out using the algorithm set forth in \citet{KASTNER2014408}, implemented in the \texttt{R}-package \texttt{stochvol}. The package moreover draws the innovation variances of the stochastic volatility processes. We use this draw to obtain the shrinkage parameters $\tau_{\sigma ij}$.
  \item Given $\{\bm{f}_t\}_{t=1}^{T}$, we simulate the factor loadings in $\bm{L}$ using standard posteriors. Conditional on a draw of the loadings, we obtain the prior variances $\tau_{Lj}$.
  \item The full history for $\{h_t\}_{t=1}^{T}$ is sampled via an independence Metropolis-Hastings algorithm \citep{10.2307/1392151}. A minor adaption required by the notion of the volatility being featured in the mean is accounted for in the respective acceptance probabilities.
  \item We update the hyperparameters $a_\bullet$ via Metropolis-Hastings steps sketched above.
\end{enumerate}
For the empirical application, we iterate this algorithm $12,000$ times and discard the initial $6,000$ draws as burn-in. We consider each third draw of the remaining $6,000$ resulting in a set of $2,000$ draws for posterior inference. It is worth mentioning that the algorithm exhibits satisfactory convergence properties.

\renewcommand\theequation{C.\arabic{equation}}
\renewcommand\thefigure{C.\arabic{figure}}
\section*{Appendix C: Homogeneity and heterogeneity across countries and over time}\label{app:addresults}
\begin{figure}[h]
\begin{subfigure}[b]{0.5\textwidth}
\caption{Time-invariant coefficients}
\includegraphics[width=\textwidth]{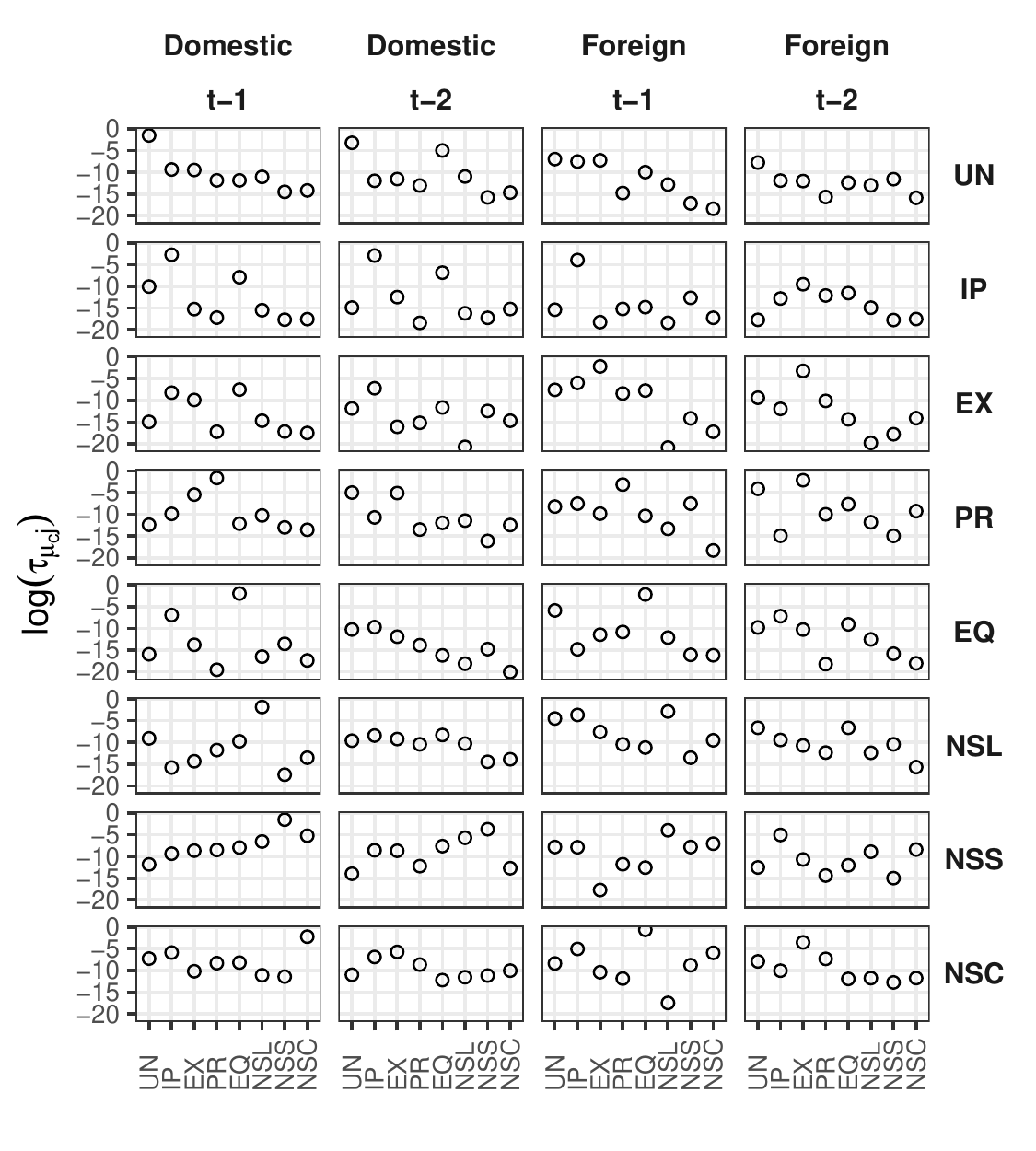}
\end{subfigure}
\begin{subfigure}[b]{0.5\textwidth}
\caption{Square root state innovation variances}
\includegraphics[width=\textwidth]{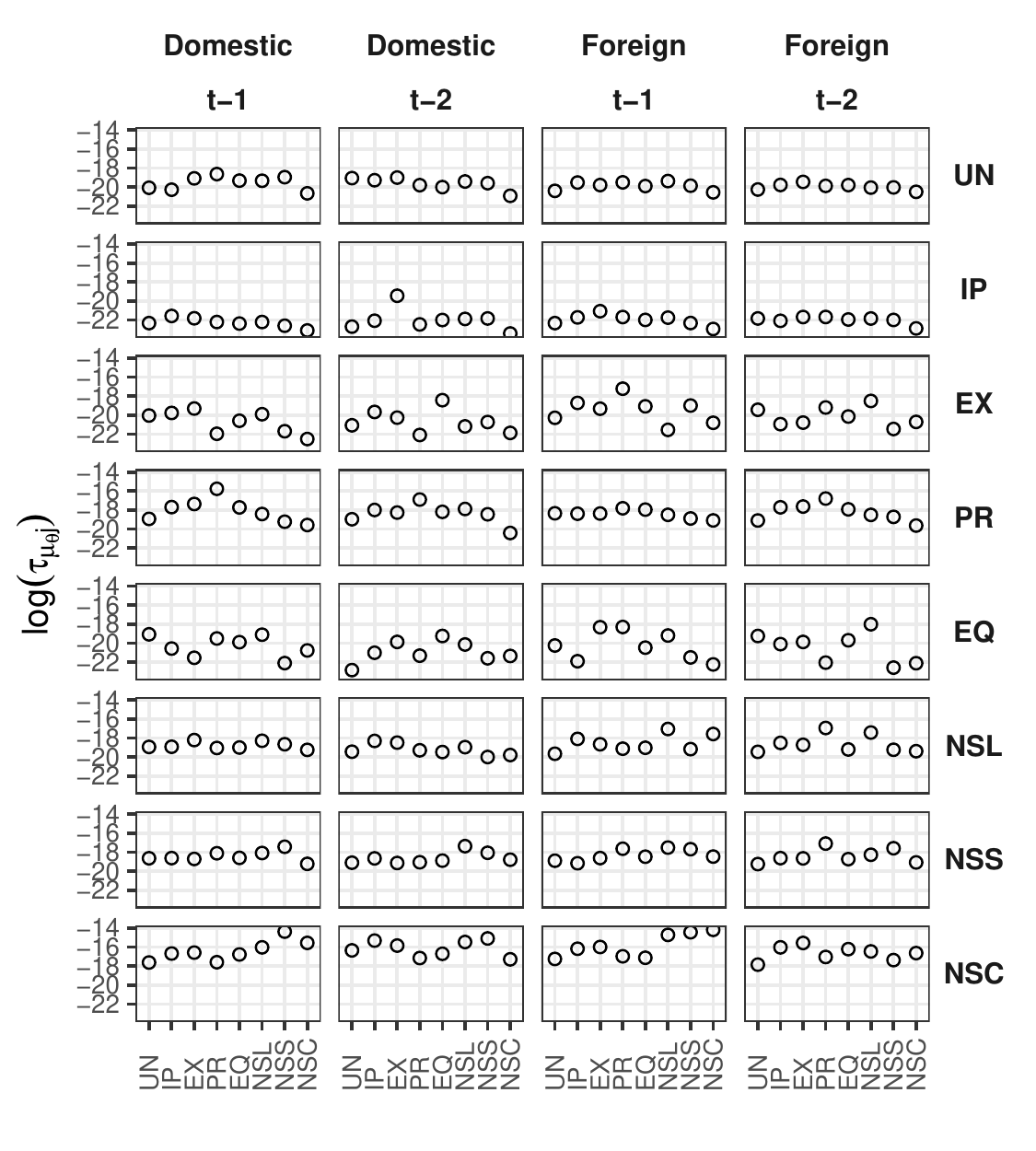}
\end{subfigure}
\caption{Log posterior mean of $\tau_{\mu_s j}$ shrinking the common mean to zero.}\label{fig:p_c0}\vspace*{-0.3cm}
\caption*{\footnotesize\textit{Note}: Columns refer to a countries' own lagged ``Domestic'' variables in $\bm{y}_{it-p}$ of lag $t-p$, while ``Foreign'' indicates the coefficients associated with $\bm{y}^{\ast}_{it-q}$ at $t-q$.}
\end{figure}

The following additional results serve to provide intuition on the properties of the prior setup. The non-centered parameterization of the state-space model allows to investigate both shrinkage on the common mean of the time-invariant part of the model coefficients $\mu_{cj}$, and the corresponding state innovation variances $\mu_{\theta j}$. Figure \ref{fig:p_c0} shows the respective posterior mean of $\tau_{\mu_c j}$ and $\tau_{\mu_\theta j}$ on the logarithmic scale. Smaller values indicate heavier shrinkage towards zero.\footnote{For visualization purposes and the imposed prior restrictions, we do not present the corresponding prior variances for the intercept term and the impact vector $\bm{\beta}_{it}$.}

The first column of Fig. \ref{fig:p_c0}(a) highlights the first own lag of each equation in $\mu_{cj}$ to feature mainly non-zero coefficients, reflected in values of $\log(\tau_{\mu_c j})$ close to zero. This implies that only little shrinkage towards zero is imposed on these coefficients by the resulting loose prior variance $\tau_{\mu_c j}$. Such patterns, albeit less distinctive, are also observable for the second lag of the domestic coefficients in the second column. However, we generally detect tighter prior variances for the second lags, with differences depending on the respective equation. Turning to the third and fourth columns that indicate shrinkage on the foreign lags per equation, we find similar shrinkage patterns when comparing to the first domestic lag. Non-domestic movements appear to play a role in the dynamic evolution of the Nelson-Siegel factors. In general, the results point towards the necessity of considering international dynamics, a feature explicitly addressed by the proposed multi-country approach.

Figure \ref{fig:p_c0}(b) provides evidence of shrinkage towards zero of the state innovation variances that drive time-variation in the model coefficients. Note, however, that shrinkage of the common mean towards zero does not necessarily imply constant model coefficients, due to additional flexibility on the second prior hierarchy. A key finding is that the unemployment and industrial production equations are pushed strongly towards a constant parameter specification both for domestic and foreign lags. A simliar picture is present in the inflation equation, albeit at a slightly lower overall degree of shrinkage induced by the respective $\tau_{\mu_\theta j}$. The higher value of $\log(\tau_{\mu_\theta j})$ on the first own domestic lag of inflation in the inflation equation suggests changes in the persistence of inflation dynamics.

\begin{figure}[t]
\begin{subfigure}[b]{0.5\textwidth}
\caption{Time-invariant coefficients}
\includegraphics[width=\textwidth]{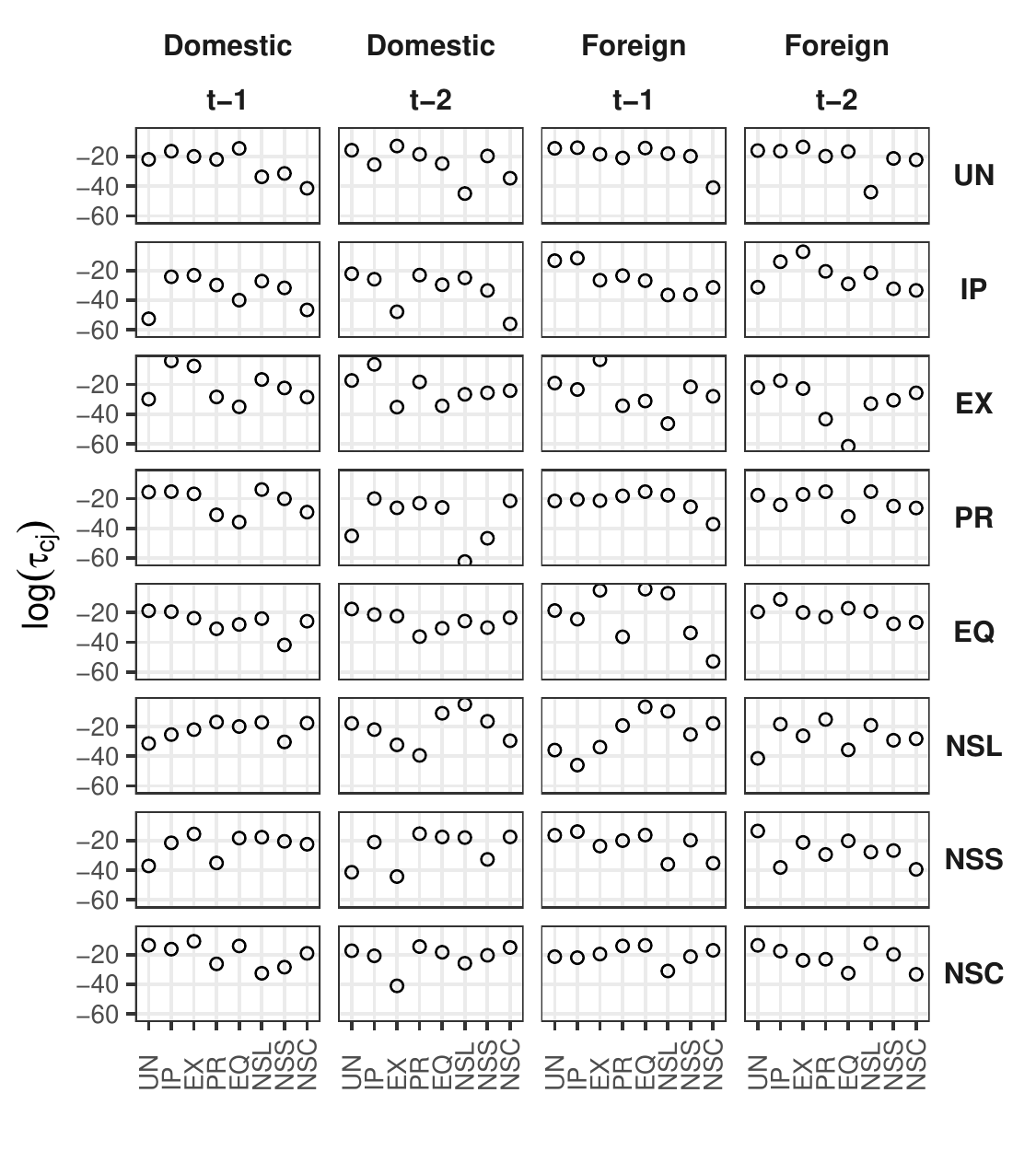}
\end{subfigure}
\begin{subfigure}[b]{0.5\textwidth}
\caption{Square root state innovation variances}
\includegraphics[width=\textwidth]{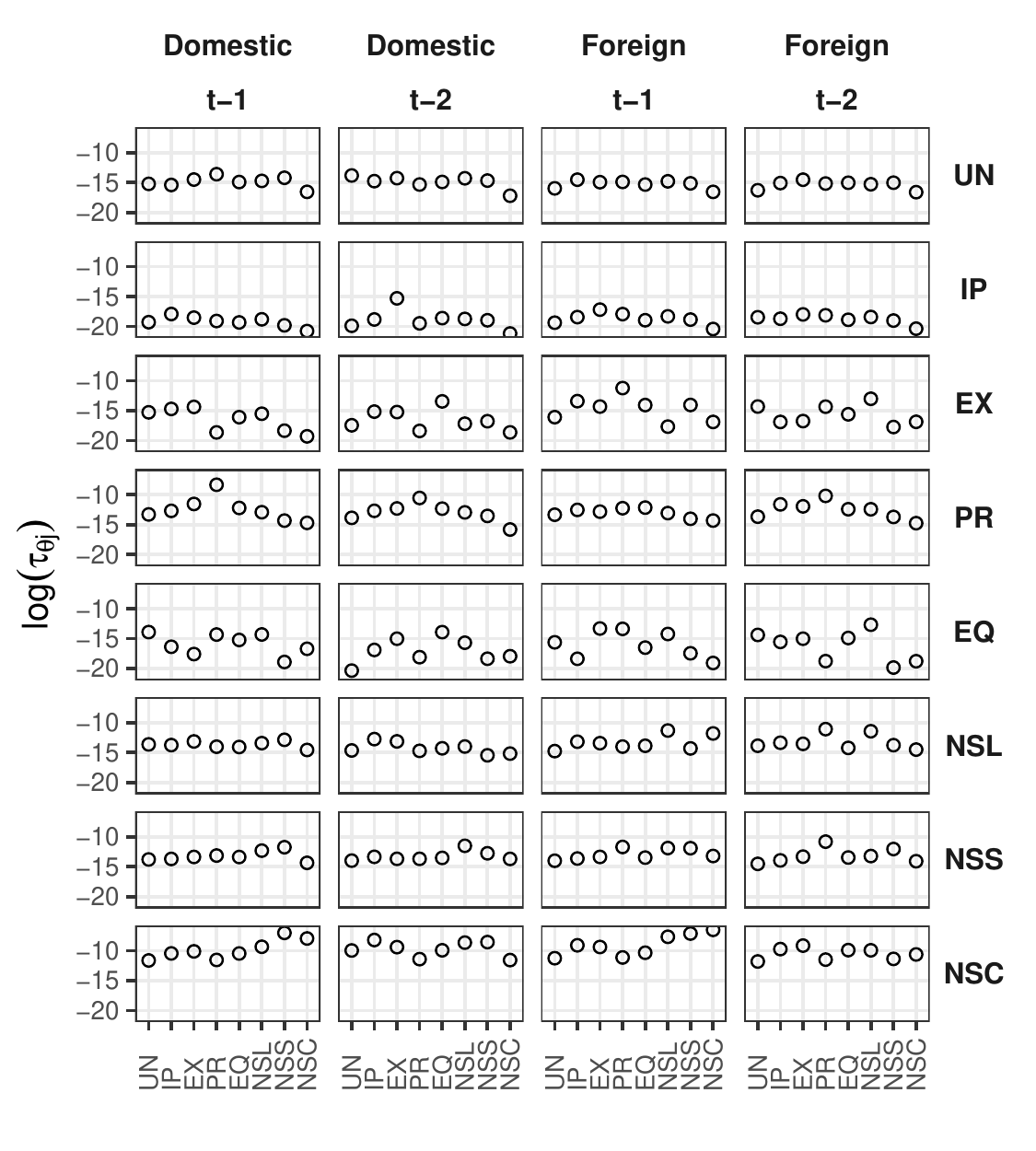}
\end{subfigure}
\caption{Log posterior mean of $\tau_{sj}$ shrinking country-specific coefficients towards $\mu_{sj}$.}\label{fig:p_V0}\vspace*{-0.3cm}
\caption*{\footnotesize\textit{Note}: Columns refer to a countries' own lagged ``Domestic'' variables in $\bm{y}_{it-p}$ of lag $t-p$, while ``Foreign'' indicates the coefficients associated with $\bm{y}^{\ast}_{it-q}$ at $t-q$.}
\end{figure}

Next, we analyze the estimated prior variances $\tau_{cj}$ and $\tau_{\theta j}$ that shrink country-specific coefficients towards $\mu_{cj}$ and $\mu_{\theta j}$, respectively. Again, we consider the posterior mean of $\log(\tau_{cj})$ and $\log(\tau_{\theta j})$ in Fig. \ref{fig:p_V0}. The scalings provide a natural measure of similarity across countries. Large negative values on the log-scale yield a situation referred to as cross-sectional homogeneity in the panel literature \citep[see][]{ecbwp:1507}.

One notable result in Fig. \ref{fig:p_V0}(a) is that all coefficients are strongly pushed towards homogeneity, suggested by predominantly large negative values for $\log(\tau_{cj})$. No clear patterns of similarities are visible across equations for both the domestic and foreign lag structure. Note that the first own domestic lags per equation usually feature less heavy shrinkage towards the common mean (except for inflation and equity prices), implying subtle differences in the persistence of the series across countries. Particularly strong evidence of homogeneity is present for subsets of domestic and foreign lags in all equations. Figure \ref{fig:p_V0}(b) displays that heavy shrinkage on the state innovation variances is applied to all domestic and foreign lags in the unemployment and industrial production equation. A similar picture emerges for the inflation equation, and the dynamics captured in the context of the Nelson-Siegel level and slope factors. However, some variables appear to require flexibility in terms of country-specific breaks in the coefficients.

Combining discussions in the context of Figs. \ref{fig:p_c0} and \ref{fig:p_V0} allows for different scenarios in terms of homogeneity across countries and the degree of sparsity: First, there is the possibility of heterogeneous non-zero coefficients and state innovation variances, in cases where both $\tau_{\mu_s j}$ and $\tau_{sj}$ are comparatively large. Here, prominent examples are provided by most first own lags of the domestic coefficients in their respective equation. Second, if both $\tau_{\mu_s j}$ and $\tau_{sj}$ are small, the prior setup implies heavy shrinkage of the country-specific parameters towards zero, for example regarding state innovation variances in the equations for unemployment and industrial production. Third, for large $\tau_{\mu_s j}$ and small $\tau_{sj}$, the prior implies homogeneous non-zero parameters featured mainly in the context of the first autoregressive foreign lags. While no clear relationship between Figs. \ref{fig:p_c0} and \ref{fig:p_V0} for the constant part of the coefficients can be identified, the adverse is true for the state innovation variances. This implies that if the common mean of the latter is non-zero on the first hierarchy of the prior, this is typically associated with less heavy shrinkage towards the common mean on the second hierarchy of the prior.

\end{appendices}
\end{document}